**My life with GianCarlo Ghirardi, a great friend, a great teacher and a great physicist.**

*Francesco de Stefano*

*Liceo Scientifico "Giovanni Marinelli"*

*Udine – Italy*

**Abstract.**

In these shorts notes, I wish to remember my relationship with GianCarlo Ghirardi, who was my supervisor in my master degree in Physics at the University of Trieste and then he became a very deep and close friend. I don't want to describe in details his scientific steps in the world of modern Physics: many others will make better than me. I want only to share which was my relationship with him as a teacher, a man and expecially as a deep friend, for more than 40 years.

*A great teacher.*

In Italy we say "Molti sono gli insegnanti, pochi i maestri" (that means, "many are the teachers, but few the masters"). GianCarlo was not simply the best teacher I had met in my life, but he was really "a master". I learned from him Quantum Mechanics (QM), the highest physical theory of the history of science. And I had the great luck to learn it directly from a physicist who has given very important and deep contributions to the conceptual foundations of QM. But which is the difference between a teacher and a master? I think that a teacher is someone who tries to transmit the culture about a specific field (Physics, Literature, Philosophy,…) and, of course, tries to make this in his best way. Doing this, he can be a great, good or bad teacher. A master is first of all a great teacher, but this is not sufficient: he/she is a teacher who makes this activity with love, trying to transmit to his/her students his/her own passion and who "walks together" the students, respecting their rythms, their own way and time to grow up. This is on the line of some of the greatest "masters" we had in Italy in the past: Maria Montessori[1] and don Lorenzo Milani[2].

GianCarlo was really a master: his lessons were always very clear and deep. He had always used a lot of time to prepare his lessons (I have seen this, when I became one of his friends), to be the clearest as possible for the students (we understand very well which is the difference between a teacher who prepares his lessons and one who "invents" at the moment what to tell!). And he has always tried to "build up" specific models (or *metaphors* as I like to say) to make clear the deepest and difficult concepts of QM. I remember again his lesson in 1977 about Heisenberg's uncertainity principle when he used a very simple and geometrical example (invented by himself) to explain it. It was 41 years ago, but his words are like "printed with the fire" in my mind. And every time I must talk about Heisenberg principle (during a lesson, as a teacher, or a public conference) I always use this model.

---

[1] Maria Montessori (1870-1952) was probably the greatest pedagogist in Italy. She was the first woman to obtain a master degree as a Medical Doctor in Italy and developed a revolutionary pedagogical method which is more applied in other countries (USA, Austria, India…) than in Italy.

[2] Don Lorenzo Milani (1923-1967) was a priest and a pedagogist whose most important writing was "Lettera a una professoressa" (Letter to a woman-teacher) in where he described a method of teaching where the teacher is "at the side of the weakest" and is a "teacher-friend".

I was one of his students and now I am a simple teacher of Mathematics and Physics in a Liceo Scientifico, but many of his students became good scientists and had a strong scientific cooperation with him (Renata Grassi, Fabio Benatti, Luca Marinatto, Angelo Bassi and others).

One day, after his retirement, he told me that the greatest thing he missed from his activity at the University was teaching, to have contacts with the young generations and trying to transmit them his experience.

I conclude this part repeating the content of a dialogue that GianCarlo had with John Stewart Bell about the importance to be a good teacher (or better "a master" in the definition I gave above). GianCarlo told me that he had this dialogue with Bell, when he came to Trieste at ICTP.

Bell:" GianCarlo, you know well how important I consider my interests in foundational problems. However I must state that I think that to devote oneself exclusively to this kind of studies is a luxury. One has also to do something more practical to get paid. This is why at CERN I am so involved in accelerator's physics"

Ghirardi: "John, you put me in a very delicate position; in the last 20 years I have worked exclusively in the field of the foundations of QM!"

Bell: "Oh no, GianCarlo! You have completely ignored that in these years, as a lecturer, you have trained entire generations of young people in teaching them quantum theory. This fact fully justifies your salary!"

And I and hundreds of students can surely confirm Bell's opinion!

*A great physicist.*

It is well known all over the world of Physics that GianCarlo was one of the most important theoretical physicists of our times. His contributions to QM are so deep and worldwide recognized that it is a pleonasm to list them. Moreover I am sure that others of his very famous colleagues will and can do it much more better than me. For this reason I will remember only a few of them and expecially those in which (for a very short time, unluckily) I was personally involved.

GianCarlo was surely one of the greatest experts about the conceptual and epistemological foundations of QM: his papers in this field are really milestones in the history of these debates. I like to underline a coincidence: GianCarlo was born in Milan in 1935. As well as all theoretical physicists konw, this was a great year in the history of QM: in 1935 Albert Einstein, Boris Podolsky and Nathan Rosen wrote a famous paper (the EPR paper, maybe the most quoted in the history of Physics) in which they challenged the so called "Copenhagen interpretation of QM", showing that QM is a correct but incomplete physical theory. In the same year, few months later, Niels Bohr replied to this paper writing exaclty the opposite: QM is complete. One month later Bohr's reply, another famous paper appeared, wrote by Erwin Schrödinger, where he presented "the actual situation in QM" and described also one of the most famous conceptual paradoxes of QM: the so called "Schrödinger's cat paradox". Ghirardi was born just in this year and it is very curious that he spent nearly 50 years of his scientific research around these topics. He was one of the most experts in the world about the EPR paper and, with his most important theory, the so called GRW theory (Ghirardi-Rimini-Weber), he solved, in a specifically dynamical way, the Schrödinger's cat paradox.

The GRW theory started with a model that the three physicists presented in 1986 in a paper whose title is "Unified dynamics for microscopic and macroscopic systems". And the title suggests very well its content: in this paper GRW described a model in order to overcome one of the greatest conceptual "puzzles" in QM: its difficulty to describe the transition to the macroscopic level from the microscopic one in a very clear dynamical way. Up to that moment, standard QM had never solved in a good way this problem (and the Schrödinger's cat paradox was developed by the austrian physicist just to underline this). I was in Trieste in GianCarlo's studio when he showed me a letter he had received from John Bell in which the Irish genius expressed his admiration about the GRW paper and underlined that he considered it one of the most important steps forward to find a solution to the problem of the transition from the micro- to the macro-level. GianCarlo was very happy to read this letter and he also felt a great responsibility, if the great John Bell considered their model in such a deep way. GianCarlo had never thought to have built (together with Rimini and Weber) a "new theory", an alternative approach to the physical world in comparison with QM: he always spoke of their "model" not "theory". But the subsequent developments of their approach (with the cooperation of Renata Grassi and Philip Pearle, for example) transformed step by step what was in the beginning a model into a real new physical theory which, together with Bohmian Mechanics, is nowadays one of the most relevant alternatives to the standard QM.

I didn't work together with him about GRW, but I want to underline that he often spoke with me about his advances in this field, asking me what was my opinion. And this is another aspect of GianCarlo's humanity: he always treated me "at the same level", just like I was able to follow his reasonings and technical progresses. But from some years I had left the academic world (unluckily, as I told before, for health reasons) and I had become a teacher in a High School and of course I couldn't really completely follow and understand what he was telling me. But I was always happy that had treated me at his same level.

And this was precisely one of his characteristics: he never considered his interlocutors "less" than him or at a "lower" level. A great physicst, but never presumptuous! And (we know this very well) this is a very rare fact.

A second important scientific contribution that I like to quote is his celebrated book about QM: "Un'occhiata alle carte di Dio", translated into English with the title "Sneaking a look at God's cards" (the title is a quotation of a famous Einstein's sentence). This book is, without any doubts, the most rigorous and, at the same time, popular introduction to the conceptual foundations of QM available in Italy. And the success of this book is the best proof. GianCarlo devoted himself for many years to write this book and I had the great privilege and honour to be chosen by him to correct the draft version. GianCarlo has always thought that common people had the right and then the pleasure to be introduced into the most important ideas of modern science. But he also knew that many popular books described, for example, QM in a too "naive" and simple way, totally avoiding the mathematical points of the theory. In the attempt to be not too difficult, these books produced very often misunderstandings about the basic concepts of QM. Then, how can be useful a book if, to be popular, pushes people to understand in a wrong way this marvellous theory? GianCarlo has decided to treat common people in an "intelligent way", writing a book which is like a beautiful excursion to a mountain: everybody who likes to practice this sport knows very well that to reach the top of a mountain is hard… sometimes very hard! But when you are on the top… what a pleasure!! You look at the things from a higher point of view and this is priceless. Then GianCarlo has decided not to avoid mathematics but to use the simplest (and at the same time the most rigorous) mathematics to obtain this result. And the book has got the point! It is not a book "to read on the beach": one must read, read again, write something, make his own calculations, but at the end he will really understand the conceptual foundations of this theory.

*A great friend.*

Now it is time to talk about GianCarlo in his best part for me: his friendship. As I told in the beginning, he was one of my teachers at the University (the best teacher or, better, master I had) and my first meeting with him was attending his very celebrated course of "History of Physics" which was in effect a course about the conceptual foundations of QM. This happened in 1976 and I was totally charmed by his way to teach and to show his passion towards QM. After the exam, I had immediately asked him if he could be my supervisor and he accepted. It is impossible for me to avoid to tell that my meeting with GianCarlo realized my dreams in the cultural field. In fact, when I was attending the Liceo Scientifico (the same school where now I am a teacher!) for some years I had not Physics in my mind. I preferred Chemistry and Biochemistry in particular. But when I was studying for the final exam ("esame di maturità", in Italian) I wrote a little paper about the atomic models and in this work I met QM! And I was so fascinating, expecially from the philosophycal features of this theory, that I decided to study Physics, in order to understand them in the correct way. And when I met GianCarlo I understood immediately that I had found the right man, and that from him I could have learned in the best way these topics. And so it was.

In 1980 I became a physicist with a thesis about Quantum Theory of Measurement and Quantum Dynamical Semigroups and then I had the possibility to start my studies of Ph.D. with him at the ICTP of Trieste. During this first year, our relationship started to be deeper and deeper. We talked about our own lives, our personal problems and we found that our souls were really very similar. I started to cooperate with him at the scientific level but this was very short. As I wrote before, during this year of Ph. D., I had problems with my health and then I couldn't succeed to give the exams that were necessary to continue. GianCarlo was very close to me in this period and I will never forget this. But I was obliged to leave the University, also because my family had not the financial resources to support me, and I started to work. Then I became a teacher beginning my career in this field. I tried to continue my cooperation with him, but it was too heavy to attend two jobs at the same time. But I remained always in contact with him: I invited him many times in Udine to make conferences about QM. And often he came without being paid: only because a friend had asked him! We have continued to talk about ourselves, our lives, our problems in life. I went to visit him many times at ICTP, I knew his wife Laura and his three daughters Monica, Barbara and Lucia. I always tried to follow him in important international conferences like those held in Urbino (1985), Cesena (1992) and many in Trieste. And during these moments I had the possibility to see other very human aspects of him. Every talk was a success, of course. But he always told me, before starting his speech, that he was not "in shape" and that probably he would have not made a good talk. On the opposite, every time he got the point! But this was GianCarlo: not a man like "I know all and now I tell you the truth!" But a man also with his weaknesses which made him not presumptuous but modest and sweet. And generous. About this I want to finish these notes with a very significant episode.

In 2007 I received a call by him and the content was nearly this: "Hallo Francesco… I want to tell you something that I hope you'll like. Do you remember the paper that you and me wrote together in 1987 and that was rejected by *Nature*? So… I and Luca Marinatto have wrote another paper and this has been accepted to be published by *Physical Review A* and I have thought to write also your name as one of the authors!" Of course I had two opposite feelings in that moment: happiness and shame! The first is obvious to explain: for the second, I was ashamed because I had not written a word of this paper! If some scientists will read these notes, they will understand very well: which scientist would make such a thing? Very few in all over the world, I'm sure! And GianCarlo was such a man.

Good bye, GianCarlo! Beyond life and death, forever friends!